# Correlation Effect on the Two-Dimensional Peierls Phase

Shutaro Chiba* and Yoshiyuki Ono*,†

*Department of Physics, Toho University, Miyama 2-2-1, Funabashi, Chiba, 274-8510, Japan
†Research Center of Materials with Integrated Properties, Toho University, Funabashi, Chiba 274-8510, Japan

**Abstract.** The effect of the electron-electron (e-e) interaction on the Peierls lattice distortions due to the electron-lattice (e-l) interaction is studied in the two-dimensional Peierls-Hubbard model, treating the fluctuation of e-e interaction around the Hartree-Fock solution within the 2nd order perturbation theory. In our previous work, using the Hartree-Fock approximation, we found multimode Peierls lattice distortions with wave vectors, the nesting vector **Q** and those parallel to it, are not affected by an e-e interaction if it is weak compared with the e-l coupling. The phase transition between the BOW (bond order wave) with multimode lattice distortions and the SDW (spin density wave) with the wave vector **Q** behaves as the 1st order transition. The property of the multimode BOW is found to change drastically when we consider the fluctuation effect within the 2nd order perturbation theory. The Fourier components of the multimode BOW increase gradually as the e-e interaction parameter is increased. Especially the Fourier component with the wave vector equal to the smallest reciprocal lattice vector in the presence of the multimode BOW is most strongly affected by the fluctuation effect.

Competition between the Peierls lattice distortion and the electron-electron (e-e) interaction is widely discussed in the 1D Peierls-Hubbard model with a half-filled electronic band. In this system the lattice distortion which has a wave number $Q = \pi$ plays a very important role in the all range of e-e interaction parameter [1].

We are interested in the same problem in the 2D system, since the ground state of the 2D electron-lattice (e-l) system, especially in the 2D SSH (Su-Schrieffer-Heeger) model with a half-filled electronic band, is an unusual Peierls state [2] in contrast to the Peierls state expected from that of the 1D system [3]. The ground state of 2D SSH model has multimode lattice distortions with the wave vector $\mathbf{Q} = (\pi, \pi)$ and those parallel to it. In addition, there are infinite number of degenerate states with non-equivalent distortion patterns corresponding to many different combinations of the Fourier components of the distortion for the wave vectors parallel to **Q** [4].

The effects of the e-e interaction on this multimode Peierls state can be discussed using the 2D Peierls-Hubbard (2D PH) model. The Hamiltonian of the 2D PH model is described as follows [5],

$$H = -\sum_{\mathbf{r},s}(t_0 - \alpha x_\mathbf{r})\left(c^\dagger_{\mathbf{r}+\mathbf{e}_x,s}c_{\mathbf{r},s} + \text{h.c.}\right)$$
$$- \sum_{\mathbf{r},s}(t_0 - \alpha y_\mathbf{r})\left(c^\dagger_{\mathbf{r}+\mathbf{e}_y,s}c_{\mathbf{r},s} + \text{h.c.}\right)$$
$$+ \frac{K}{2}\sum_\mathbf{r}(x_\mathbf{r}^2 + y_\mathbf{r}^2) + U\sum_\mathbf{r} n_{\mathbf{r},\uparrow}n_{\mathbf{r},\downarrow}, \quad (1)$$

where $x_\mathbf{r} \equiv u^x_{\mathbf{r}+\mathbf{e}_x} - u^x_\mathbf{r}$, $y_\mathbf{r} \equiv u^y_{\mathbf{r}+\mathbf{e}_y} - u^y_\mathbf{r}$ with $\{u^{x(y)}_\mathbf{r}\}$ and $\mathbf{e}_{x(y)}$ the lattice displacements at the site **r** and the unit lattice vector for respective direction, respectively, and the parameters $t_0$, $\alpha$, $K$ and $U$ are the electron transfer integral between nearest-neighbor sites, the e-l coupling constant, the force constant describing the ionic coupling strength and the strength of the on-site e-e interaction between up and down spins, respectively.

In our previous paper [6] we used the Hartree-Fock approximation (HFA) as the easiest approach to discuss the e-e interaction. In the HFA, the multimode lattice distortions are not affected by the e-e interaction, and at the critical value of $U$, which depends on the dimensionless e-l coupling constant $\lambda \equiv \frac{\alpha^2}{Kt_0}$, we found the 1st order phase transition between the multimode BOW (bond order wave) and the antiferromagnetic SDW (spin density wave).

It is important to discuss the e-e interaction effect on the multimode Peierls state using more accurate treatment of the e-e interaction than the HFA. In this paper we discuss the 2nd order fluctuation effect due to the e-e interaction around the HFA on the 2D Peierls phase. Omitting the expression for the e-e interaction Hamiltonian within the HFA, we show here the fluctuation part, which is ignored in the HFA,

$$H^{\text{flu}}_{\text{e-e}} = U\sum_\mathbf{r}\left(n_{\mathbf{r},\uparrow} - \langle n_{\mathbf{r},\uparrow}\rangle\right)\left(n_{\mathbf{r},\downarrow} - \langle n_{\mathbf{r},\downarrow}\rangle\right). \quad (2)$$

The lattice distortion at the site **r** can be calculated by the following equation,

$$x^{[j]}_\mathbf{r}(y^{[j]}_\mathbf{r}) = -\frac{\alpha}{K}\int_{-\infty}^\infty \frac{d\omega}{2\pi i} G^{[j]}_{\mathbf{r},\mathbf{r}+\mathbf{e}_{x(y)}}(\omega). \quad (3)$$

where $x^{[j]}$ and $y^{[j]}$ are the $j$-th order correction of the lattice distortion obtained by treating eq. (2) up to the $j$-th order perturbation, and $G^{[j]}_{\mathbf{r},\mathbf{r}+\mathbf{e}_a}(\omega)$ ($a=x,y$) is the $j$-th order Green function. Note that $j=0$ in eq. (3) represents the lattice distortion when we treat the e-e interaction within the HFA.

The lattice distortion as a function of $U$ is represented in Fig. 1 in terms of the absolute values of the Fourier components $x_q$ and $y_q$. We assume the system size to be $L \times L = 8 \times 8$ and $12 \times 12$, the e-l coupling constant to be $\lambda = 0.5$ and the distortion being composed of the Fourier components for $\mathbf{Q}$ and $\mathbf{Q}/2$ which is one of the simplest distortion patterns among many degenerate multimode Peierls states [4]. In the multimode Peierls state, it is known that $|x_q| = |y_q|$. In Fig. 1 (a), the data for the $\mathbf{Q}$ component are shown for $L = 8$ ($\times$) and 12 (+) while Fig. 1 (b) represents the data for the $\mathbf{Q}/2$ component. The results obtained within the HFA are drawn for reference.

As will be seen in Fig. 1, the amplitude of the $\mathbf{Q}$ component decreases with increasing $U$ in the case of $L = 8$ but increases in the case of $L = 12$, while the $\mathbf{Q}/2$ component increases with $U$ in both cases. We have confirmed the similar behavior of the $U$ dependence in the case with $L = 16$ as in the case with $L = 12$. Presumably, the system size appears more strongly in the case of small size. Hereafter, therefore, we discuss the results for the case with $L = 12$. The $\mathbf{Q}/2$ component is enhanced more rapidly with increasing $U$ than the $\mathbf{Q}$ component, and the former gets dominant as $U$ approaches the critical value $U_c = 2.15 t_0$ at which the 1st order phase transition from the multimode BOW to the SDW with no lattice distortion occurs. Similar enhancement of the lattice distortion is known also in the 1D PH model, where the lattice distortion with the wave number $Q(=\pi)$ is increased with increasing $U$ as far as $U$ is smaller than the band width ($\simeq 2t_0$) [1]; when $U$ exceeds the band width, the Fourier component of the lattice distortion decreases gradually but remains finite even in the limit $U \gg t_0$. The increase of the lattice distortion results in the suppression of the charge fluctuation until $U$ reaches the band width. In our calculation the similar behavior of the lattice distortion is confirmed. Since it is argued in the 1D case that the enhancement of the lattice distortion is due to the Umklapp process involved in the fluctuation effect, it will be plausible to consider that the most affected mode of the lattice distortion is the one with the reciprocal lattice vector in the presence of the distortion, which is equal to $\mathbf{Q}/2$ in the present case. The data shown in Fig. 1 are consistent with this scenario.

The phase transition between the multimode BOW and the SDW is essentially the same as that determined within the HFA and it behaves as the 1st order transition even when we consider electronic fluctuations within the second order perturbation. It is possible that the order of the phase transition and the critical value of $U$ change if we treat the e-e interaction more rigorous way. However, if the enhancement of the lattice distortion continues to the intermediate values of $U$, we may expect that the spin-Peierls state with multiple Fourier components of lattice distortions occurs. It is well known that in the limit of $U \gg t_0$ the 2D PH model can be mapped on to the 2D Heisenberg model in which the effect of lattice distortions is included in the spin exchange interaction. We have studied this model by the exact diagonalization for small clusters, and found the multimode spin-Peierls phase [7]. We will discuss this 2D multimode spin-Peierls state elsewhere.

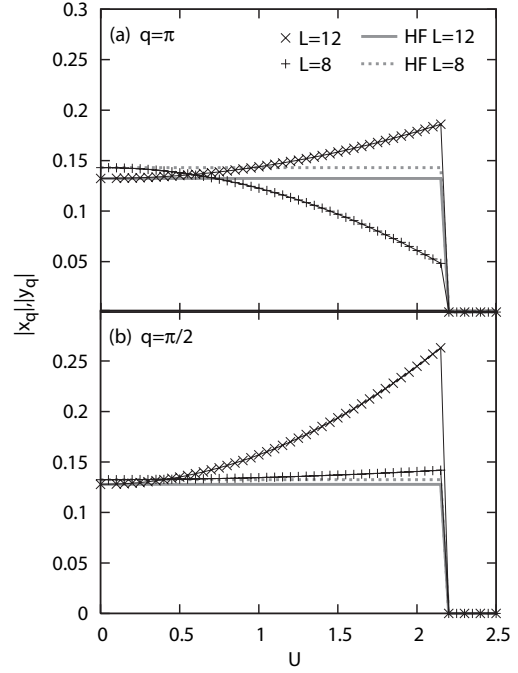

**FIGURE 1.** The amplitudes of the Fourier components $x_q$ and $y_q$ of the lattice distortion in unit of the lattice constant are plotted as functions of $U$ (scaled by $t_0$), for $L = 8$ ($\times$) and 12 (+) in the case with $\lambda = 0.5$. Note that, in the multimode Peierls state, $|x_q| = |y_q|$. The distortion consists of the components with $\mathbf{Q}$ (a) and $\mathbf{Q}/2$ (b). The data obtained within the HFA [6] are shown by the solid ($L = 12$) and the dotted ($L = 8$) line.